\documentclass[runningheads]{cl2emult}

\usepackage{makeidx}  
\usepackage{graphicx} 
\usepackage{subeqnar} 
\usepackage{multicol} 
\usepackage{cropmark} 
\usepackage{eso}      
\makeindex            

\begin{document}
\title*{Reconstructing the microwave sky using a combined
maximum-entropy and mexican hat wavelet analysis}
%
%
\toctitle{Reconstructing the microwave sky using a combined
maximum-entropy and mexican hat wavelet analysis}
%
%
\titlerunning{Combined MEM and MHW analysis of CMB observations}
%
\author{R.~Bel\'en Barreiro\inst{1}
\and Patricio Vielva\inst{2,3}
\and Michael~P.~Hobson\inst{1}
\and Enrique Mart\'\i nez-Gonz\'alez\inst{2}
\and Anthony~N.~Lasenby\inst{1}
\and Jos\'e~L.~Sanz\inst{2}
\and Luigi Toffolatti\inst{4}}
\authorrunning{R.Bel\'en Barreiro et al.}
%
%
\institute{Astrophysics Group, Cavendish Laboratory, Madingley Road,\\
Cambridge CB3 0HE, UK
\and Instituto de F{\'\i}sica de Cantabria (CSIC -- UC),
Fac. Ciencias, \\
Avda. de los Castros s/n, 39005 Santander, Spain
\and 
Departamento de F{\'\i}sica Moderna, Universidad de Cantabria, \\
Avda. de los Castros s/n, 39005 Santander, Spain
\and Departamento de F{\'\i}sica, Universidad de Oviedo, c/ Calvo Sotelo
s/n, \\ 33007 Oviedo, Spain}

\maketitle              

\begin{abstract}
We present a combined maximum-entropy method (MEM) and Mexican Hat
wavelet (MHW) analysis in order to recover the different
components of the microwave sky. We apply this technique to simulated
observations by the ESA Planck satellite in small patches of the
sky. In particular, the introduction of the MHW allows one to detect
and subtract the brightest point sources present in the input data and
therefore to improve the reconstructions of the CMB and foreground
components achieved by MEM on its own. In addition, a
point source catalogue at each Planck frequency is produced, which is
more complete and accurate than those obtained by each technique
independently. 
\end{abstract}

\section{Introduction}
Cosmic Microwave Background (CMB) observations carry a wealth of
information about the Universe. Indeed, an accurate knowledge of the
CMB anisotropies can place tight constraints on fundamental parameters
as well as to discriminate between competing theories of structure
formation. Future CMB experiments such as  
the NASA MAP satellite and the Planck mission from ESA, will
provide with multifrequency data at high resolution and
sensitivity. 
However, these data contain not only the cosmological signal but also 
Galactic foregrounds, extragalactic point sources,
thermal and kinetic Sunyaev-Zelodvich (SZ) emission from cluster of
galaxies and instrumental noise. Therefore our capacity to recover all
the valuable information encoded in the CMB
will critically depend on our ability to denoise and separate the cosmological
signal from the rest of components of the microwave sky.

To perform such a separation,~\cite{h98} has developed a Fourier 
MEM algorithm. This technique is particularly successful at using
multifrequency data to identify foreground emission from physical
components whose spectral signatures are (reasonably) well-known.
Therefore, the most problematic foreground to remove is that due to
extragalactic point sources, since each source has a unique frequency spectrum
and, moreover, is notoriously difficult to predict.
To adress this problem,~\cite{h99} extended the MEM approach to deal with
point sources as an extra `noise' contribution.
A different approach was developed by~\cite{sanz}
who showed that the Mexican Hat wavelet (MHW) is in
fact the optimal pseudo-filter for detecting point sources under
reasonable conditions. The application of this wavelet to realistic
simulations was presented in~\cite{c00} and extended in~\cite{v00}.

The aim of this work is to show that the MEM and MHW techniques are
actually complementary and can be combined to improve the accuracy of
the separation of diffuse foregrounds from the CMB and increase the
number of point sources that are identified and successfully subtracted.
The joint analysis has been performed on simulated observations by the
Planck satellite but it could be
straightforwardly applied to other multifrequency CMB experiments such
as the forthcoming NASA MAP satellite or the recently performed
Boomerang and MAXIMA experiments.

\section{The MEM and MHW joint analysis}
A detailed description of the MEM algorithm is given in~\cite{h98}
and~\cite{h99} whereas the MHW method is explained in~\cite{c00} and
~\cite{v00}. Therefore, we will
focus on how the two approaches can be successfully combined to
produce a more powerful joint analysis scheme (see also~\cite{mexmem}).

The MEM technique presented in~\cite{h99} includes point sources as
part of a generalised noise vector. This has proved to be very successful
at performing a full component separation with the contamination due
to point sources greatly reduced in the reconstructions. 
Moreover, by comparing the input data maps with `mock' data
obtained from the separated components it is possible to obtain point
source catalogues at each observing frequency.
Since point sources are modelled as an additional noise,
MEM performs well in identifying and removing a large number of point
sources with low to intermediate fluxes. However, it is rather poorer
at removing the contributions from the brightest point sources. These
tend to remain in the reconstructed maps, 
although with significantly reduced amplitudes.

The MHW technique is based in the fact that the point sources are very much
amplified with respect to the background in the wavelet coefficients
map. Therefore, the detection of point sources is performed in
wavelet space at a certain optimal scale instead of in real space.
This method out-performs, in general, other tecnhiques such as
SExtractor (\cite{bertin}) and standard harmonic filtering
(\cite{tegm}). In addition, this method
does not require any assumptions to be made regarding the
statistical properties of the point source population or the
underlying emission from the CMB (or other foreground components).
The MHW is particularly
efficient in detecting the brightest point sources. Moreover, their
amplitude is also accurately estimated. For weaker sources, however,
the MHW performs more poorly by either inaccurately estimating the
flux or failing to detect the source altogether.

The strenght and weakness of the MEM and MHW approaches clearly
indicate that they are complementary and that a combined analysis
might lead to improve results as compared to using each method on its
own. Therefore we propose the following technique for analysing
multifrequency observations of the CMB that contain point source
contamination. First, the MHW is applied at each 
observing frequency map and the brightest point sources detected and
subtracted. 
The processed data maps are then used as input for the MEM algorithm
in order to 
perform a separation of the physical components as explained 
in~\cite{h99}. This leads to more accurate reconstructed maps, mostly free from
point source contamination. These reconstructions
are then used to generate `mock' data, which are subtracted from the
input data to generate data residuals maps at each observing
frequency. Since the diffuse components are
reasonably well recovered, these residuals maps will mostly
contain point sources and instrumental noise. Finally the MHW is
applied on each of these maps in order to recover a more complete and
accurate catalogue than those obtained by each technique independently.

\section{Foreground separation}
We have applied the MEM and MHW joint technique to simulated observations of
the Planck satellite in small patches of the sky ($12.8^\circ \times
12.8^\circ$). Our simulated data
contain a Gaussian CDM model for the CMB with $\Omega=0.3$ and
$\Omega_\Lambda=0.7$ for which the power spectrum was generated using
CMBFAST (\cite{cmbfast}). They also include
thermal and kinetic SZ effects (following the model of~\cite{diego}), 
Galactic foregrounds 
(synchrotron, dust and free-free), extragalactic point sources
(simulated according to~\cite{toff}) and
instrumental noise at the level expected in the Planck data. 
A description of these simulations as well as the
observational parameters used for the Planck satellite are given
in~\cite{mexmem}.
Fig.~\ref{inputs} shows the input maps for the CMB, kinetic and thermal
SZ effects and Galactic foregrounds.
\begin{figure}
\centering
\caption[]{The $12.8\times 12.8$ deg$^2$ realisations of the six input
components used to produce the simulated Planck data. The different
panels correspond to (from left to right and from top to bottom) CMB,
kinetic SZ effect, thermal SZ effect, Galactic dust,
Galactic free-free and Galactic synchrotron emission. 
Each component is plotted at 300 GHz and has been convolved with a
Gaussian beam of FWHM 5 arcmin (the highest resolution expected for the Planck
satellite). The map units are equivalent thermodynamic temperature in $\mu$K}
\label{inputs}
\end{figure}
In order to perform the separation and reconstruction of the different
components we have assume knowledge of the azimuthally averaged power
spectrum of these six input components (see~\cite{h98} for more
details). Using the model of ~\cite{toff} we
have also introduced the power spectrum of the point sources at each
frequency channel, including cross power spectra between
channels. 
However, the recovery of the main components and point sources do not
depend critically on this assumption (see ~\cite{mexmem})

The resulting reconstructions of the physical components at a
reference frequency of 300~GHz are shown in Fig.~\ref{rec}. 
We see that the main input components have been faithfully
recovered and no obvious visible contamination of point sources remain
in the reconstructions. We give the rms reconstruction errors for each
component in Table~\ref{errors}.
For comparison, the rms error of the reconstructed maps without a
previous subtraction of point sources using the MHW is also given.
\begin{figure}
\centering
\caption[]{As in Fig.~\ref{inputs} but for the reconstructed maps}
\label{rec}
\end{figure}
\begin{table}
\centering
\caption{The rms in $\mu$K of the reconstruction residuals smoothed 
with a 5 arcmin FWHM Gaussian beam with and without the initial
subtraction of bright point sources using the MHW.
For comparison the rms of the input maps 
are also given} 
\renewcommand{\arraystretch}{1.}
\setlength\tabcolsep{3pt}
\begin{tabular}{cccccccccc}
\hline\noalign{\smallskip}
Component  & input & error & error \\
& rms & (with MHW) & (without MHW) \\
\hline
CMB        & 112.3 & 7.68 & 8.62 \\
Kinetic SZ & 0.69 & 0.70 & 0.70 \\
Thermal SZ & 5.37 & 4.64 & 4.66 \\
Dust       & 55.8 & 2.68 & 3.39 \\
Free-Free  & 0.66 & 0.22 & 0.24 \\
Synchrotron& 0.32 & 0.11 & 0.12 \\
\end{tabular}
\label{errors}
\end{table}
In particular, the reconstruction of the CMB map is very good, 
with a rms reconstruction error of 7.7$\mu$K which corresponds to an
accuracy of $\sim$6.8 per cent level as compared to the rms of the
input map. Even more impressive is the reconstruction of the dust
map. Although the high frequency channels, where the dust is the main
component, are highly contaminated by infrared sources, none of them
are visible in the dust reconstructed map. The main features of the
free-free emission are also recovered mostly due to its
high correlation with the dust. The reconstructed synchrotron map is
basicaly a lower resolution image of the input. This is expectable
since the only channels that provide useful information 
about this component are the lowest frequency ones which also have the
lowest angular resolutions.
Regarding the reconstruction of the thermal SZ, most of the bright
clusters have been reproduced and only a few point sources have been
misidentified as clusters. At the reference frequency of reconstruction
these point sources appear mostly as negative features.
Finally, as expected, the reconstruction of the kinetic SZ is quite
poor and only a few clusters whose corresponding thermal SZ is large
have been detected.


%
\begin{table}
\centering
\caption{The point source catalogues obtained using the MHW alone
(MHWc), MEM alone (MEMc) and the joint analysis method (M\&Mc).
For each Planck observing frequency, we list the number of detected 
sources, the flux limit of the catalogue and the mean percentage
error for the amplitude estimation}
\renewcommand{\arraystretch}{1}
\setlength\tabcolsep{3pt}
\begin{tabular}{cccccccccc}
\hline\noalign{\smallskip}
& & MHWc & & & MEMc & & & M\&Mc & \\
\hline
Freq. &
No. & Min Flux & ${E}_{abs}$ &
No. & Min Flux & ${E}_{abs}$ &
No. & Min Flux & ${E}_{abs}$ \\
(GHz) &
detect. & (Jy) & (\%) &
detect. & (Jy) & (\%) &
detect. & (Jy) & (\%) \\
\hline
30 & 4 & 0.46 & 12.1 &
	21 & 0.10 & 12.1 &
	19 & 0.10 & 12.3 \\
44 & 3 & 0.58 & 6.7 &
	11 & 0.24 & 8.6 &
	11 & 0.24 & 8.4 \\
70 & 5 & 0.28 & 21.0 &
	19 & 0.12 & 10.5 &
	18 & 0.15 & 8.1 \\
100 (L) & 3 & 0.59 & 6.8 &
	16 & 0.13 & 15.6 &
	14 & 0.13 & 10.0 \\
100 (H) & 7 & 0.27 & 7.7 &
	33 & 0.08 & 13.4 &
	33 & 0.07 & 12.9 \\
143 & 4 & 0.40 & 13.6 &
	1 & 0.10 & 14.2 &
	8 & 0.06 & 24.2 \\
217 & 5 & 0.25 & 9.9 &
	1 & 0.10 & 23.4 &
	8 & 0.06  & 19.0 \\
353 & 10 & 0.07 & 34.6 &
	6 & 0.24 & 41.4 &
	9 & 0.24 & 8.2 \\
545 & 29 & 0.26 & 20.4 &
	13 & 0.23 & 39.4 &
	41 & 0.24 & 14.4 \\
857 & 86 & 0.58 & 10.4 &
	107 & 0.41 & 17.6 &
	150 & 0.31 & 13.8 \\
\hline
\end{tabular}
\label{cat}
\end{table}
\section{Point source catalogues}

A main aim of the Planck mission is also to produce accurate point
source catalogues at each of the observing frequencies. In this
section we will focus on how the combination of the MEM and MHW techniques
can improve the catalogues obtained by each of them
independently. 

The MHW catalogue (MHWc) is produced in the way explained in ~\cite{v00}.
The MEM and joint analysis catalogues (MEMc and M\&Mc) are constructed
applying the MHW to the data residuals maps obtained as explained in $\S2$.
Table~\ref{cat} gives the number of point sources, minimum flux
reached and average error in the estimation of the amplitude for the
three catalogues.

In the low frequency channels, the MEM and M\&M catalogues contain a
similar number of point sources, being much more complete than
the MHWc. In this case, the contribution of the MHW is only
to improve the amplitude estimation of a few bright point sources.
The MHW can only detect a few point sources in these channels
due to the large beam size, what means that CMB and point sources
have a similar characteristic scale.

The improvement is far more noticeable in the intermediate and high
frequency channels. On the one hand, a larger number of point sources
are detected with the combined technique than with each of the methods
independently. This is due to the complementary nature of the two
approaches, so that bright sources are detected by the MHW whereas
fainter sources are identified by MEM. On the other hand, the
amplitude of the point sources is more accurately estimated in the
M\&Mc. When MEM is used without previously applied the MHW, the bright
point sources, which are not well characterised by a generalised noise, 
tend to remain in the reconstructions. This produces a bias in the
estimation of the amplitude of the bright point sources which are
underestimated. This problem is solved with the combined technique.
We also point out that, although the average error in the amplitude
estimation can be higher in the M\&Mc due to the detection of a larger
number of faint point sources, those point sources present in all
three catalogues are, in average, better estimated with the joint analysis.

\section{Conclusions}
We have presented a combined analysis of the maximum-entropy method
and the Mexican Hat wavelet to separate and reconstruct the physical
components of the microwave sky from multifrequency observations of
the CMB that contain point sources. 
We have applied this technique to simulated data of the
Planck satellite pointing out the improvements achieved due to the
complementary nature of both approaches. 
Bright point sources are identified and subtracted by 
the MHW whereas MEM is able to deal with
fainter point sources as a generalised noise.
As a result, the reconstructions of the CMB, SZ
effects and Galactic foregrounds are improved and mostly free of
contaminating point sources. Moreover, using the joint analysis 
more complete and accurate
point source catalogues are produced at each observing frequency as
compared to those obtained by each of the techniques independently.

\clearpage
\addcontentsline{toc}{section}{Index}
\flushbottom
\printindex

\end{document}